\documentclass[figures]{epl}
\usepackage{amsmath}
\usepackage{amssymb}

\title{Relating monomer to centre-of-mass distribution functions in
polymer solutions}
\shorttitle{Distribution functions in polymer solutions}
\author{V. Krakoviack \and J. P. Hansen \and A. A. Louis}
\shortauthor{V. Krakoviack \etal}
\institute{Department of Chemistry, University of Cambridge,
Lensfield Road, Cambridge CB2 1EW, United Kingdom}


\begin{document}

\maketitle

\begin{abstract}
A relationship between the measurable monomer-monomer structure
factor, and the centre-of-mass (CM) structure factor of dilute or
semi-dilute polymer solutions is derived from Ornstein-Zernike
relations within the ``polymer reference interaction site model''
(PRISM) formalism, by considering the CM of each polymer as an
auxiliary site and neglecting direct correlations between the latter
and the CM and monomers of neighbouring polymers. The predictions
agree well with Monte Carlo data for self-avoiding walk polymers, and
are considerably more accurate than the predictions of simple
factorization approximations.
\end{abstract}

There have recently been a number of attempts to represent polymer
coils in solution as systems of soft, penetrable particles of fixed
\cite{bolhuis2001,pagonabarraga2001} or variable \cite{murat1998}
shape, and size proportional to the radius of gyration $R_g$, which
interact via effective pair forces obtained by averaging over
individual monomer degrees of freedom. A similar approach has proved
very successful in the description of star polymer solutions
\cite{likos1998}. The advantage of this reductionist strategy is that
the effective pair forces act only between single interaction sites
within each polymer, rather than between the large number of
individual monomers or Kuhn segments belonging to the interpenetrating
coils. A natural (but by no means unique) choice of the single
interaction site is the centre-of-mass (CM) of the polymers. This
coarse-graining leads to an enormous reduction (by a factor equal to
the number $\nu$ of monomers or segments in each polymer) in the
number of interacting degrees of freedom, thus allowing, \textit{inter
alia}, efficient simulation of large scale phenomena involving many
polymers.

The price to pay is that the effective interactions are
state-dependent, \textit{i.e.} the pair potentials resulting from the
coarse-graining procedure depend in general on polymer concentration,
temperature, and the degree of polymerization, although the latter
dependence is negligible for sufficiently large $\nu$ such that the
scaling regime is reached in practice. In ref.~\cite{bolhuis2001} the
state-dependent effective pair potential between the CM of
non-intersecting self-avoiding walk (SAW) polymers in dilute and
semi-dilute solutions was derived from a very accurate Statistical
Mechanics inversion procedure of the CM pair distribution function
$g_{cc}(r)$. The latter had to be determined for each polymer
concentration from fully microscopic Monte Carlo (MC) simulations of
samples involving hundreds of SAW polymers. This ``exact'' procedure
is obviously very computer intensive, and in a certain sense defeats
the original purpose, which is to replace a detailed, monomer level
description of polymer coils by a projected representation involving
only their CM's.

To overcome this methodological bottleneck, one is naturally led to
the use of approximate theories of intermolecular correlations between
polymers. One obvious candidate is the ``polymer reference interaction
site model'' (PRISM), which provides a successful theory of such
correlations, particularly so in the melt, where polymers are known to
behave as Gaussian coils \cite{schweizer1997}. However PRISM provides
only monomer-monomer pair distribution functions $g_{mm}(r)$, and an
accurate procedure is needed to extract the CM-CM pair distribution
function $g_{cc}(r)$ from a knowledge of $g_{mm}(r)$. In this Letter
the PRISM formalism is extended to derive such a relationship, which
also involves the form factors (or internal structure factors) of
individual polymers. The relation is tested against MC simulation
data, and turns out to be vastly superior to previous proposals
\cite{koyama1979,pagonabarraga2001}.

Consider a monodisperse solution of $N$ linear polymers with $\nu$
monomers or segments each, in a volume $V$; the polymer number density
is $\rho=N/V$. Let $\mathbf{R}^i$ ($1\leq i\leq N$) be the CM position
vector of the $i$th polymer, and
$\mathbf{r}^i_\alpha=\mathbf{R}^i+\mathbf{u}^i_\alpha$
($1\leq\alpha\leq\nu$) the positions of the $\nu$ monomers in that
chain. The Fourier components of the monomer and CM density operators
are:
\begin{subequations}
\begin{gather}
\rho_\mathbf{q}=\sum_{i=1}^N \sum_{\alpha=1}^\nu 
e^{i\mathbf{q}\mathbf{r}_\alpha^i}\,,\\
\rho^{\text{CM}}_\mathbf{q}=\sum_{i=1}^N e^{i\mathbf{q}\mathbf{R}^i}\,.
\end{gather}
It proves also convenient to define the Fourier components of the
intramolecular monomer density of the $i$th polymer:
\begin{equation}
\rho^i_\mathbf{q}=\sum_{\alpha=1}^\nu
e^{i\mathbf{q}\mathbf{u}_\alpha^i}\,,
\end{equation}
\end{subequations}
in terms of which one may express the following form factors, which
characterize the internal structure of each individual polymer coil,
in the presence of all surrounding polymers:
\begin{subequations}
\begin{gather}
\omega_{mm}(q)=\frac{1}{\nu} <\rho^i_\mathbf{q}
\rho^i_\mathbf{-q}>\,,\label{mmffact}\\
\omega_{cm}(q)=<\rho^i_\mathbf{q}>\,.\label{cmffact}
\end{gather}
\end{subequations}
The brackets denote canonical averages weighted by the Boltzmann
factor involving the total interaction energy between all monomers on
all polymers, and accounting for the connectivity constraints. The
resulting form factors are independent of the index $i$, because all
polymers are equivalent, and depend only on the modulus
$q=|\mathbf{q}|$ of the wave vector due to rotational invariance (the
polymer solutions are isotropic).

The total monomer structure factor, as measured e.g. by coherent
neutron or light scattering experiments, is defined by
\begin{equation}
S_{mm}(q)=\frac{1}{N\nu^2}<\rho_\mathbf{q}\rho_{-\mathbf{q}}>
=\frac{1}{N\nu^2}<\sum_i \sum_j \sum_\alpha \sum_\beta 
e^{i\mathbf{q}(\mathbf{r}_\alpha^i-\mathbf{r}_\beta^j)}>\,,
\end{equation}
and naturally splits into intramolecular ($i=j$) and intermolecular
($i\neq j$) contributions:
\begin{equation}\label{interintra}
S_{mm}(q)=S_{mm}^{\text{intra}}(q)+S_{mm}^{\text{inter}}(q)=
\frac{\omega_{mm}(q)}{\nu}+S_{mm}^{\text{inter}}(q)\,.
\end{equation}
The intermolecular contribution may be rewritten as:
\begin{equation}\label{predecoup}
S_{mm}^{\text{inter}}(q)=\frac{1}{N\nu^2}\sum_i\sum_{j\neq i}
<e^{i\mathbf{q}(\mathbf{R}^i-\mathbf{R}^j)}
\rho^i_\mathbf{q}\rho^j_\mathbf{-q}>\,.
\end{equation}
A common decoupling approximation is to assume that the intramolecular
conformations of any two polymers are independent of each other and of
the mutual positions of their CM, so that the statistical average in
\eqref{predecoup} factorizes according to:
\begin{equation}\label{postdecoup}
S_{mm}^{\text{inter}}(q)\simeq\frac{1}{N\nu^2}\sum_i \sum_{j\neq i}
<e^{i\mathbf{q}(\mathbf{R}^i-\mathbf{R}^j)}><\rho^i_\mathbf{q}>
<\rho^j_\mathbf{-q}>
=\frac{\omega_{cm}(q)^2}{\nu^2}[S_{cc}(q)-1]\,,
\end{equation}
where $S_{cc}(q)$ is the CM structure factor:
\begin{equation}\label{comsfact}
S_{cc}(q)=\frac{1}{N}<\rho^{\text{CM}}_\mathbf{q}
\rho^{\text{CM}}_{-\mathbf{q}}>=1+\rho h_{cc}(q)\,,
\end{equation}
and $h_{cc}(q)$ is the Fourier transform of the CM-CM pair correlation
function $h_{cc}(r)=g_{cc}(r)-1$. Substituting eqs.~\eqref{postdecoup}
and \eqref{comsfact} into eq.~\eqref{interintra}, one arrives at
\begin{equation}
S_{mm}(q)=\frac{\omega_{mm}(q)}{\nu}+\rho h_{mm}(q)
=\frac{\omega_{mm}(q)}{\nu}+\rho \frac{\omega_{cm}(q)^2}{\nu^2}
h_{cc}(q)\,,
\end{equation}
which leads to the desired approximate relation between the Fourier
transforms of the monomer-monomer and CM-CM correlation functions:
\begin{equation}
h_{cc}(q)=\frac{\nu^2}{\omega_{cm}(q)^2}h_{mm}(q)\,.\label{koyama}
\end{equation}
Equation~\eqref{koyama} is the approximation proposed by Koyama
\cite{koyama1979} and is formally identical to the so-called
``free-rotation'' approximation for rigid molecules; it requires a
knowledge of the form factor $\omega_{cm}(q)$ defined by
eq.~\eqref{cmffact}.

An alternative approximation is based on the rigid particle assumption
used by Pagonabarraga and Cates \cite{pagonabarraga2001}, according to
which intramolecular conformations and CM positions are independent,
and intramolecular conformations of different polymers are correlated
as if they belonged to the same polymer (``rigid particle''
assumption), \textit{i.e.}:
\begin{align}
S_{mm}(q)\simeq&\frac{1}{N\nu^2}\sum_i \sum_j
<e^{i\mathbf{q}(\mathbf{R}^i-\mathbf{R}^j)}>
<\rho^i_\mathbf{q}\rho^j_\mathbf{-q}>\nonumber\\
=&S_{cc}(q)\frac{1}{\nu^2}<\rho^i_\mathbf{q}\rho^i_\mathbf{-q}>\nonumber\\
=&\frac{\omega_{mm}(q)}{\nu}S_{cc}(q)\,,
\end{align}
leading to
\begin{equation}
h_{cc}(q)=\frac{\nu}{\omega_{mm}(q)}h_{mm}(q)\,.\label{pagona}
\end{equation}

The relations \eqref{koyama} and \eqref{pagona} are clearly based on
uncontrolled factorization approximations; they will be tested below
against ``exact'' simulation data, and compared to the relation which
we now set out to establish within the framework of the PRISM theory
\cite{schweizer1997}. PRISM is based on the assumption that all
correlation functions between monomers are independent of their
positions along the chain, \textit{i.e.} end effects are neglected,
which is true in the scaling limit $\nu\to\infty$. This unique
correlation function $h_{mm}(q)$ is related to a unique
monomer-monomer direct correlation function $c_{mm}(q)$ by the PRISM
Orstein-Zernike (OZ) relation:
\begin{equation}\label{prismoz}
h_{mm}(q)=\omega_{mm}(q)c_{mm}(q)[\omega_{mm}(q)+\rho\nu h_{mm}(q)]\,.
\end{equation}

The key idea now is to consider the CM of each polymer as an
additional non-interacting site, which is linked to the monomer
position vectors by the defining ``connectivity'' constraint
\begin{equation}
\mathbf{R}^i=\frac{1}{\nu}\sum_{\alpha=1}^\nu \mathbf{r}_\alpha^i\,.
\end{equation}
This trick of introducing auxiliary sites to compute special
correlation functions dates back to the early days of the ``reference
interaction site model'' (RISM) theory (from which PRISM is an
extension) \cite{chandler1972} and has been first proposed by Chandler
\cite{chandler1973}. Since the CM auxiliary site does not interact
with any of the $\nu$ segments, it must clearly be treated separately;
thus, each polymer now has two species of sites, namely the CM and the
$\nu$ equivalent interaction sites associated with the physical
segments.

The single OZ relation \eqref{prismoz} is now replaced by a $2\times
2$ matrix of OZ relations. The latter are further simplified by the
plausible assumption that the direct correlation functions between the
CM of one polymer, and the CM as well as the interaction sites of the
other polymers are identically zero, \textit{i.e.}
\begin{subequations}\label{closure}
\begin{gather}
c_{cc}(q)\equiv 0\,,\label{closurecc}\\
c_{mc}(q)=c_{cm}(q)\equiv 0\,.
\end{gather}
\end{subequations}
The four coupled OZ relations are then given by \eqref{prismoz}, which
is not modified by the presence of the auxiliary, non-interacting site
\cite{lue1995}, together with
\begin{subequations}
\begin{gather}
h_{cm}(q)=\omega_{cm}(q)c_{mm}(q)[\omega_{mm}(q)+\rho\nu h_{mm}(q)]\,,\\
h_{mc}(q)=\omega_{mm}(q)c_{mm}(q)[\omega_{mc}(q)+\rho\nu h_{mc}(q)]\,,\\
h_{cc}(q)=\omega_{cm}(q)c_{mm}(q)[\omega_{cm}(q)+\rho\nu h_{cm}(q)]\,.
\end{gather}
\end{subequations}

These equations immediately lead to the desired relation:
\begin{equation}\label{main}
\ h_{cc}(q)=\frac{\omega_{cm}(q)^2}{\omega_{mm}(q)^2}h_{mm}(q)\,,
\end{equation}
which is independent of any specific closure relation, except for the
assumptions \eqref{closure}.

\begin{figure}
\onefigure[angle=270]{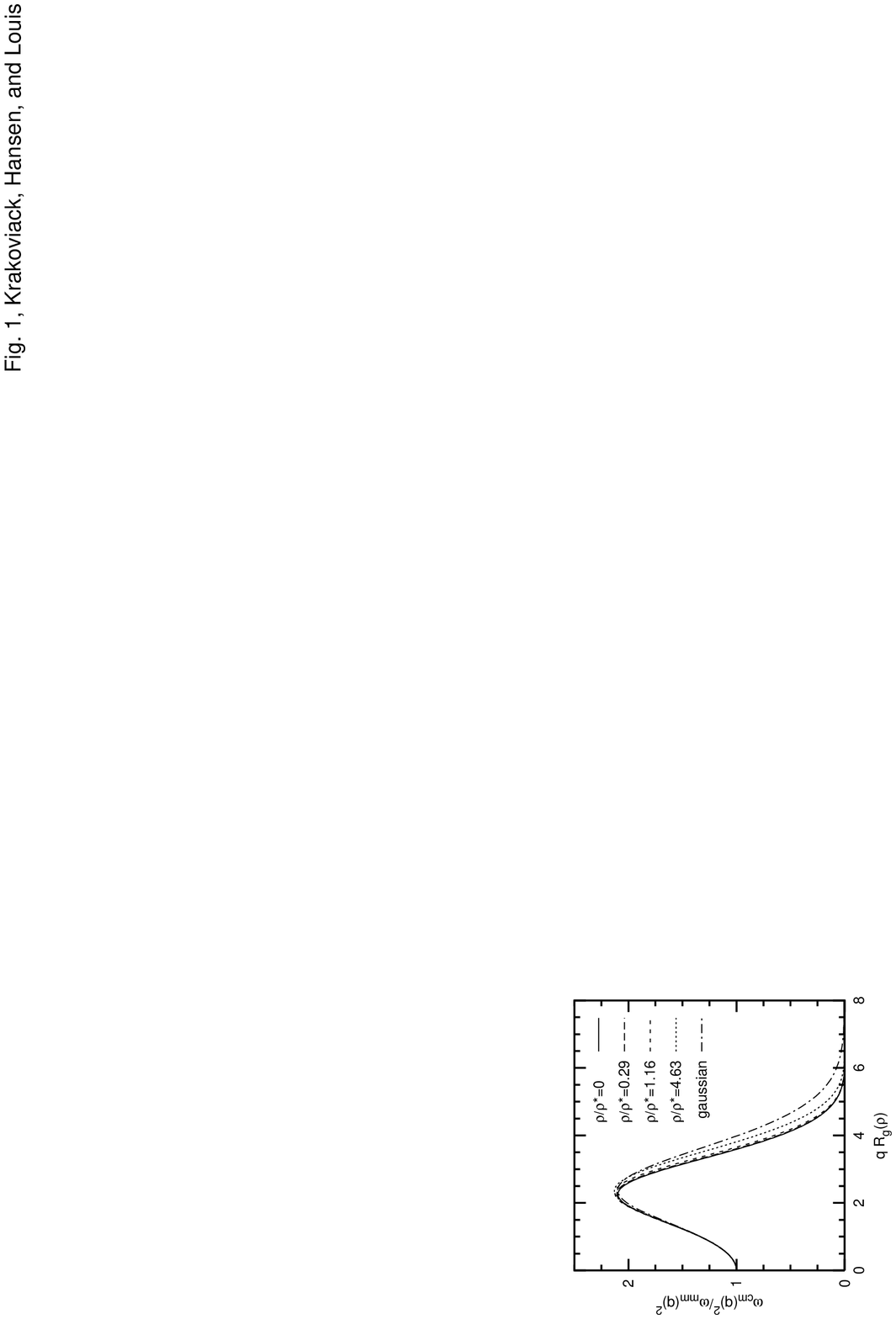}
\caption{\label{fig1} Density dependence of the ratio
$\omega_{cm}(q)^2/\omega_{mm}(q)^2$ as a function of the wave number
scaled by the density-dependent radius of gyration $R_g(\rho)$, as
obtained by MC simulations of $\nu=500$ SAW polymers on a simple cubic
lattice. For comparison, the corresponding function for Gaussian
chains is also shown.}
\end{figure}

Equation \eqref{main} provides the desired link between the
monomer-monomer correlation function, which may be calculated from
PRISM theory, and the CM-CM correlation function which is the basic
input into the coarse-graining scheme, since the state dependent
effective pair potential between the centres of mass follows directly
from standard inversion procedures \cite{bolhuis2001}. To extract
$h_{cc}(q)$ from $h_{mm}(q)$ also requires a knowledge of the form
factors $\omega_{mm}(q)$ and $\omega_{cm}(q)$. These are known
explicitly for Gaussian chains \cite{doi}, and PRISM theory generally
uses the Gaussian chain form factor in the OZ relation
\eqref{prismoz}, which is strictly valid only in the melt. In polymer
solutions, we expect the form factors to depend on polymer
concentration. To investigate this dependence, we have carried out MC
simulations of $\nu=500$ SAW polymers on a simple cubic lattice for
reduced densities $\rho/\rho^*=0$, $0.29$, $1.16$, and $4.63$, where
the overlapping density $\rho^*$ is defined through the relation
$4\pi\rho^*R_g^3(\rho=0)/3=1$. Results for the ratio
$\omega_{cm}(q)^2/\omega_{mm}(q)^2$, relevant for the mapping
\eqref{main} of $h_{mm}(q)$ onto $h_{cc}(q)$, are shown in
fig.~\ref{fig1}. They are compared with the corresponding ratio for
Gaussian chains, which is found to qualitatively reproduce the shape
of the curves obtained by computer simulation. When the wave number
$q$ is scaled with the radius of gyration $R_g(\rho)$ appropriate for
each density, which decreases from 16.8 to 14.8 lattice spacings as
the density increases, the low density results ($\rho/\rho^* \lesssim
1$) are seen to fall practically on a single master curve. The results
at $\rho/\rho^*=4.63$, which is well into the semi-dilute regime,
deviate from this master curve, towards the result for a Gaussian
chain, as one might expect.

\begin{figure}
\twofigures[angle=270]{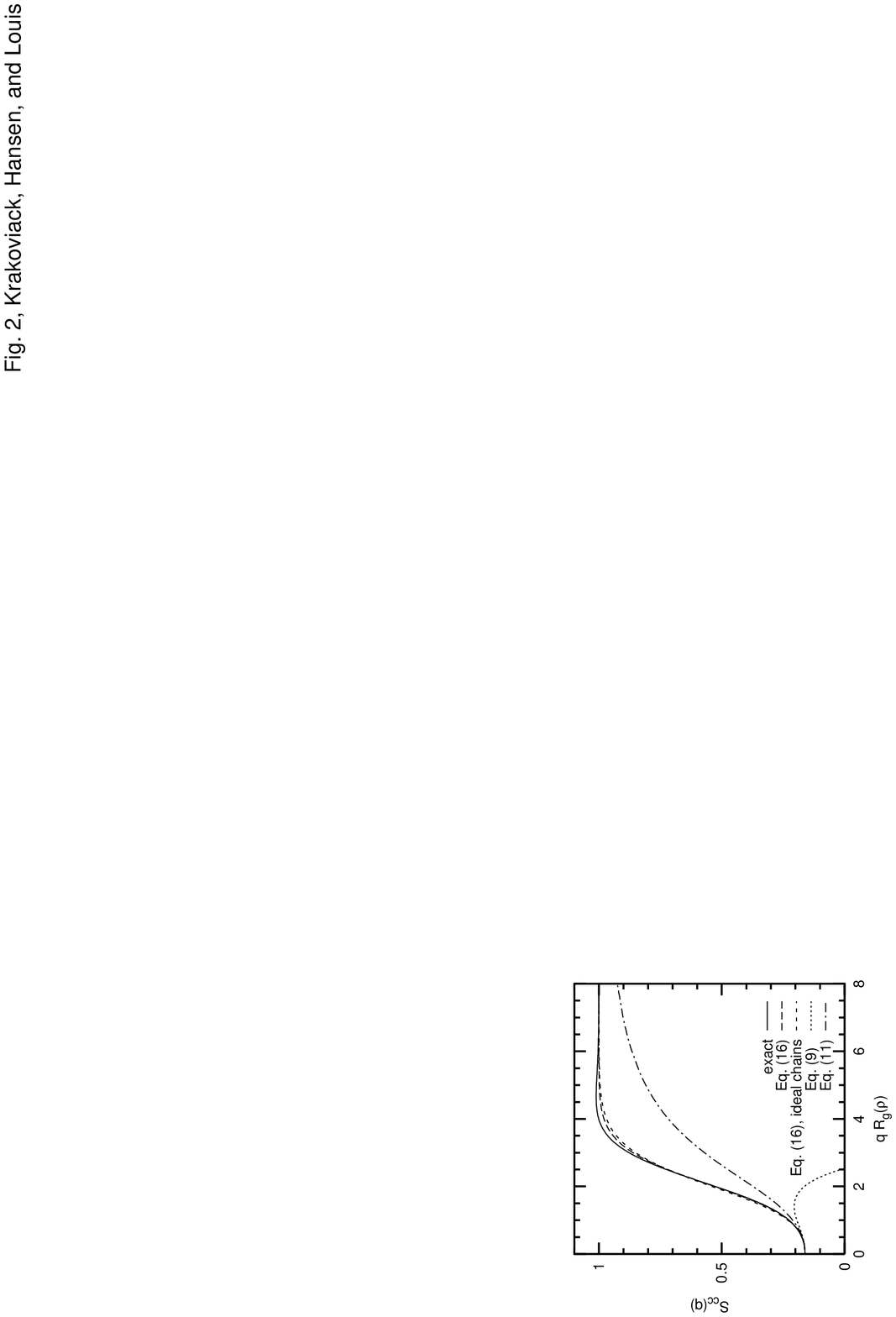}{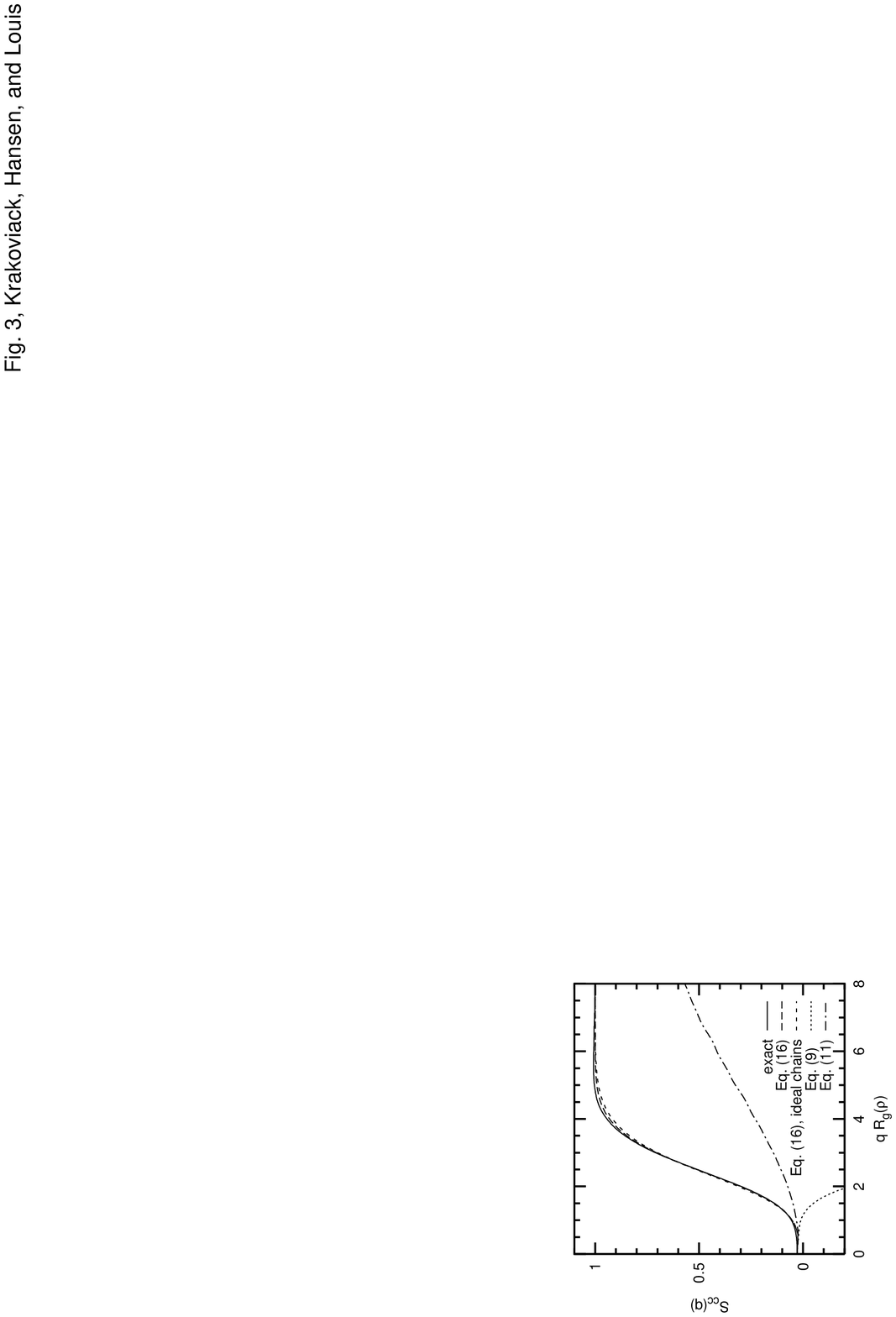}
\caption{\label{fig2} CM structure factor $S_{cc}(q)$ for $\nu=500$
SAW polymers on a simple cubic lattice at a reduced density
$\rho/\rho^*=1.16$. The ``exact'' structure factor is compared with
the various approximations proposed in the text using the ``exact''
monomer correlation function $h_{mm}(q)$ and form factors
$\omega_{cm}(q)$ and $\omega_{mm}(q)$. For comparison, in the case of
approximation \eqref{main}, results based on the Gaussian chain form
factors are also shown.}
\caption{\label{fig3} Same as fig.~\ref{fig2}, at a reduced density
$\rho/\rho^*=4.63$.}
\end{figure}

Having determined the ``exact'' form factors, we are now in a position
to test the relations between $h_{mm}(q)$ and $h_{cc}(q)$, as
predicted by the factorization approximations \eqref{koyama} and
\eqref{pagona}, and by the PRISM-based equation \eqref{main}. For the
test to be meaningful, we have used as input the ``exact'' (rather
than PRISM-generated) $h_{mm}(q)$. Figure~\ref{fig2} shows the MC
results for $S_{cc}(q)$, together with the results obtained by
combining eq.~\eqref{comsfact} with each of the three approximations
\eqref{koyama}, \eqref{pagona} and \eqref{main}, using the ``exact''
form factors, at $\rho/\rho^*=1.16$. In the case of approximation
\eqref{main}, results based on the Gaussian chain form factors are
also shown. The agreement between the ``exact'' structure factor and
the predictions from eq.~\eqref{main} is seen to be excellent at small
$q$; some deviations are seen at intermediate wave numbers $qR_g\simeq
4$, where the ``exact'' $S_{cc}(q)$ has a small maximum. Similar
agreement is found at the other densities, as shown in fig.~\ref{fig3}
for $\rho/\rho^*=4.63$. The decoupling approximations \eqref{koyama}
and \eqref{pagona} are seen to fail. Interestingly, approximation
\eqref{main} combined with the analytic Gaussian form factors also
yields good agreement.

The success of eq.~\eqref{main} in obtaining the CM correlations from
the monomer correlations suggests that the inverse route could also be
profitable. This may be expected to be more difficult because one
appears to move from less information (CM) to more information
(monomers). In fact, attempts to invert eq.~\eqref{main} result in
problems of a rather general kind because $\omega_{cm}(q)$, which can
be zero, is now in the denominator, leading to spurious singularities.
$\omega_{cm}(q)$ for Gaussian coils does not cross zero, so that it
may be used more profitably to obtain $S_{mm}(q)$ from $S_{cc}(q)$,
but the results are still not as good as what is seen in
figs.~\ref{fig2} and \ref{fig3}. We found empirically that the inverse
of eq.~\eqref{main} still works well for $q R_g \leq 2$, but for
larger $q$ deviations occur.  While the other two approximations do
not show the spurious singularity, they do show other spurious
effects.

To conclude, we have proposed a new relation between the
monomer-monomer and CM-CM correlation functions of polymers based on
an extension of PRISM theory. It has been tested for polymers in good
solvent and it is found to be quantitatively accurate except for
discrepancies at intermediate wave numbers where $h_{cc}(q)$ changes
sign and exhibits a small maximum, while $h_{mm}(q)$ is a
monotonically increasing function of $q$. This discrepancies must be
traced back to assumptions \eqref{closure}, which are the only
approximations involved here (apart from the neglect of end effects in
PRISM, that is known to have minor consequences). The factorization
approximations \eqref{koyama} and \eqref{pagona}, on the other hand,
appear to perform less well.  It should be added that approximation
\eqref{pagona} was derived for use in the melt, a regime which has not
been tested in this paper, and where the connection between CM and
monomer correlations may be different.

The present analysis stresses the need for developing a reliable
theory for the state-dependent form factors $\omega_{mm}(q)$ and
$\omega_{cm}(q)$, in particular in the dilute and semi-dilute solution
regimes that have been investigated here. In this respect, it is worth
stressing that $\omega_{cm}(q)$ has hitherto attracted very little
attention \cite{schafer1988}, probably because it is not of direct
experimental relevance. Our MC results point to a near universality of
the ratio $\omega_{cm}(q)/\omega_{mm}(q)$ as a function of $qR_g$ in
the dilute regime. Accurate form factors in the dilute and semi-dilute
regime would also provide a crucial input into PRISM calculations of
$h_{mm}(q)$. Work along these lines is in progress.

Various extensions of this work are currently under
investigation. Firstly, using the two-component PRISM OZ relation
\cite{fuchs2000}, it is easy to extend the present formalism to
colloid-polymer mixtures and simple inhomogeneous situations (polymers
near a wall for instance), that have recently attracted much
attention. Secondly, the relation \eqref{main} does not rely on any
assumption on the monomer interactions, so it would be interesting to
test its relevance for other classes of polymers such as linear
polymers in $\theta$ or poor solvents, branched polymers,
polyelectrolytes, etc. This could be very helpful in interpreting
experiments, which typically only have access to monomer-monomer
correlations.

\acknowledgments

The authors are grateful to Ludger Harnau for valuable suggestions.
VK acknowledges support from the EPSRC under grant number GR$/$M88839
and AAL acknowledges support from the Isaac Newton Trust, Cambridge.


\begin{thebibliography}{99}
\bibitem{bolhuis2001} \Name{Louis A. A., Bolhuis P. G., Hansen J. P. \and
Meijer E. J.} \Review{Phys. Rev. Lett.} \Vol{85} \Year{2000}
\Page{2522}; \Name{Bolhuis P. G., Louis A. A., Hansen J. P. \and
Meijer E. J.} \Review{J. Chem. Phys.} \Vol{114} \Year{2001}
\Page{4296}.
\bibitem{pagonabarraga2001} \Name{Pagonabarraga I. \and Cates M. E.}
\Review{Europhys. Lett.} \Vol{55} \Year{2001} \Page{348}.
\bibitem{murat1998} \Name{Murat M. \and Kremer M.}
\Review{J. Chem. Phys.} \Vol{108} \Year{1998} \Page{4340}.
\bibitem{likos1998} \Name{Likos C. N., L\"owen H., Watzlawek M., Abbas
B., Jucknischke O., Allgaier J. \and Richter D.}
\Review{Phys. Rev. Lett.} \Vol{80} \Year{1998} \Page{4450}.
\bibitem{schweizer1997} For a review, see \Name{Schweizer K. S. \and
Curro J. G.} \Review{Adv. Chem. Phys.} \Vol{98} \Year{1997} \Page{1}.
\bibitem{koyama1979} \Name{Koyama R.} \Review{Makromol. Chem.}
\Vol{181} \Year {1980} \Page{1987}; \Review{Macromolecules} \Vol{14}
\Year{1981} \Page{1299}.
\bibitem{chandler1972} \Name{Chandler D. \and Andersen H. C.}
\Review{J. Chem. Phys.} \Vol{57} \Year{1972} \Page{1930}.
\bibitem{chandler1973} \Name{Chandler D.} \Review{J. Chem. Phys.}
\Vol{59} \Year{1973} \Page{2742}.
\bibitem{lue1995} Analogous results in the framework of RISM theory
can be found in \Name{Lue L. \and Blankschtein D.}
\Review{J. Chem. Phys.} \Vol{102} \Year{1995} \Page{5460}.
\bibitem{doi} \Name{Doi M. \and Edwards S. F.} \Book{The Theory of Polymer
Dynamics} \Publ{Clarendon, Oxford} \Year{1986}.
\bibitem{schafer1988} \Name{Sch\"afer L. \and Kr\"uger B.}
\Review{J. Phys. France} \Vol{49} \Year{1988} \Page{749}.
\bibitem{fuchs2000} \Name{Fuchs M. \and Schweizer K. S.}
\Review{Europhys. Lett.} \Vol{51} \Year{2000} \Page{621};
\Review{Phys. Rev. E} \Vol{64} \Year{2001} \Page{021514}.
\end{thebibliography}
\end{document}